\documentclass[aps,prb,  twocolumn,showpacs,preprintnumbers,groupedaddress]{revtex4}

\usepackage{graphicx}
\usepackage{dcolumn}
\usepackage{bm}

\begin{document}


\title{ Effect of suppression of local distortion  on magnetic, electrical and thermal transport  properties of Cr substituted bi-layer manganite LaSr$_{2}$Mn$_{2}$O$_{7}$}

\author{M.Matsukawa} 
\email{matsukawa@iwate-u.ac.jp }
\author{M.Chiba}
\author{E.Kikuchi}
\affiliation{Department of Materials Science and Technology, Iwate University , Morioka 020-8551 , Japan }
\author{R.Suryanarayanan}
\author{M.Apostu}
\author{}
\affiliation{Laboratoire de Physico-Chimie de L'Etat Solide,CNRS,UMR8648
 Universite Paris-Sud, 91405 Orsay,France}
\author{}
\affiliation{}
\author{S.Nimori}
\affiliation{National Institute for Materials Science, Tsukuba 305-0047 ,Japan}
\author{K. Sugimoto}
\affiliation{X-Ray Research Laboratory, Rigaku Corporation, Tokyo 196-8666,Japan}


\date{\today}

\begin{abstract}
We have investigated magnetic, electrical and thermal transport properties  
(Seebeck effect and thermal conductivity) of    LaSr$_{2}$Mn$_{2-y}$Cr$_{y}$O$_{7}$  
polycrystalline samples ($y$=0.1, 0.2, 0.4 and 0.6).  The Cr$^{3+}$ substitution for Mn$^{3+}$ sites causes  a removal  of  $d_{x^2-y^2}$ orbital  of  $e_g$-electron resulting in a volume shrinkage of lattice. Magnetic measurements reveal  the appearance of   a glassy behavior for Cr-doped samples,   accompanied by both a collapse of the A-type antiferromagnetic structure and the growth of ferromagnetic clusters.  
Cr-doping effect on electrical transport strongly enhances an insulating behavior over a wide range of temperature,  while it suppresses a local minimum of thermoelectric power  at lower temperatures. For all polycrystalline samples with Cr-substitution, the variable-range-hopping (VRH) conduction model gives a reasonable fit to both resistivities and Seebeck coefficients.   
The phonon thermal conduction gradually rises with increasing Cr content, which is in contrast to a typical impurity effect on thermal conductivity. We attribute this to a suppression of local lattice distortion through the introduction of  Jahn-Teller  inactive  ions of Cr$^{3+}$.

\end{abstract}

\renewcommand{\figurename}{Fig.}
\maketitle
\section{ INTRODUCTION}
The discovery of colossal magnetoresistance (CMR)  effect in doped manganites with perovskite 
structure has stimulated considerable interest for the understanding of their physical properties \cite{TO00}. Though the insulator to metal (IM) transition and its associated CMR are well explained  on the basis of  the double exchange (DE) model,  it is pointed out that the dynamic Jahn-Teller (JT) effect due to the strong electron-phonon interaction, plays a significant role in 
the appearance of CMR as well as the DE interaction \cite{ZE51,MI95}.  Furthermore, Dagotto 
et al propose a phase separation model where  the ferromagnetic (FM) 
metallic and antiferromagnetic (AFM) insulating clusters coexist and their model strongly
 supports recent experimental studies on the physics of manganites  \cite{DA01,DA03}. 

 In bilayer manganites La$_{2-2x}$Sr$_{1+2x}$Mn$_{2}$O$_{7}$, in which a MnO$_{2}$ bilayer is alternatively stacked with a (La,Sr) $_{2}$O$_{2}$ blocking layer along the $c$-axis,  the physical properties strongly depend on hole doping level, $x$ \cite{HI98}. In particular, neutron diffraction study on half doped LaSr$_{2}$Mn$_{2}$O$_{7}$($x$=0.5) has revealed the coexistence of the A-type antiferromagnetic (AFM) phase  and CE-type antiferromagnetic charge-ordered/orbital-ordered (CO/OO) phase\cite{KU99}.   It is well known that  the CE-type CO/OO state in cubic manganites  is unstable against Cr-substitution for Mn-site and  lightly Cr doping up to a few percents yields a drastic collapse of the CO/OO phase, resulting in a FM metallic phase even in the absence of any applied magnetic  field \cite{RA97,KI99}.   While several reports on the effect of Cr substitution on the physical properties of the cubic manganites have appeared ,  very few reports have appeared on such studies in the case of bilayer manganites\cite{GU99,CH02}.   
Here,  we give some comments on pressure effect on a two-dimensional network of MnO$_6$ octahedra  in bilayer mangantes La$_{1.2}$Sr$_{1.8}$Mn$_{2}$O$_{7}$.  Argyriou et al \cite{AR97}., reported that the Mn-O(3)-Mn bond angle is almost unchanged by application of pressure, indicating no tilting of the MnO$_6$ octahedra  in the $ab$ plane.  Thus,  it is possible to examine the internal and external pressure effect in bilayered manganites, varying the bond length of the MnO$_6$ octahedra  but keeping the bond angle almost $180^ {\circ}$. 
In this paper, we report  magnetic, electrical and thermal transport properties of  single-phase  LaSr$_{2}$Mn$_{2-y}$Cr$_{y}$O$_{7}$ polycrystalline samples ($y$=0.1, 0.2, 0.4 and 0.6).  The Cr-substitution for Mn sites causes a monotonic shrink of $a(b)$-axis in contrast with a gradual elongation of $c$-axis, accompanied by $d_{x^2-y^2}$ orbital deficiencies of $e_g$-electron as listed in Table\ \ref{table1} .  The 3$d$ electronic state of  Cr$^{3+}$ ion is taken as $t_{2g}^3e_{g}^0$ (spin quantum number $S$=3/2) ,  resulting in undistorted CrO$_6$ octahedron sites free from local Jahn-Teller effect.  This finding is quite reasonable with a volume shrinkage observed due to Cr-doping because  a removal  of  $d_{x^2-y^2}$ orbital from Mn$^{3+}$ sites easily causes a suppression of local lattice distortion as discussed later.  In the parent material LaSr$_{2}$Mn$_{2}$O$_{7}$, a majority phase of  the A-type AFM state coexists with a minority phase of CE-type AFM charg-ordered/orbital-ordered  state \cite{KU99}. 
We focus our attention on Cr-doping effect  on the A-type AFM majority phase because it is expected that the CO/OO minority phase is strongly suppressed by Cr-doping. 
\section{EXPERIMENT}

Polycrystalline samples of  LaSr$_{2}$Mn$_{2-y}$Cr$_{y}$O$_{7}$ ($y$=0.1, 0.2, 0.4 and 0.6) were synthesized  by solid-state reaction of La$_{2}$O$_{3}$, SrCO$_{3}$, MnO$_{2}$ and Cr$_{2}$O$_{3}$ powders with high purity. 
The oxygen concentration of typical samples with y=0.2 and 0.6 was determined using the infrared absorption method because the existence of the Cr ions may affect the valence estimation of Mn ion made by the chemical analysis.  
The composition of cations was examined  by inductively coupled plasma analysis. 
 For the y=0.2 and 0.6 samples, we got  La$_{1.02}$Sr$_{2.01}$Mn$_{1.83}$Cr$_{0.19}$O$_{6.99}$ at $y=0.2$ and La$_{1.01}$Sr$_{1.95}$Mn$_{1.41}$Cr$_{0.59}$O$_{7.05}$ at $y=0.6$. 
Thus, we conclude that our samples prepared by the solid state reaction technique are close to nominal compositions. Let us consider the difference in oxygen concentration (hole concentration).
The values of  7-$\delta$= 6.99 at y=0.2 and 7-$\delta$=7.05 at y=0.6 give  hole contents of $x$=~0.5 and $x$=0.55,respectively. Recent neutron powder diffraction studies on La$_{2-2x}$Sr$_{1+2x}$Mn$_{2}$O$_{7}$ revealed the magnetic and crystallographic phase diagram in the region $x>0.5$  \cite{LI00}. 
 In particular, when $0.42<x<0.66$, appears the  A-type AFI state with antiferromagnetic coupling along the $c$-axis between FM single layers within one bilayer.  We believe that the excess oxygen content (x=0.55) gives little effect on the magnetic property because the AFM magnetic transition temperature is stable over the range of hole concentration up to $x=0.6$ .  

\begin{figure}[ht]
\includegraphics[width=8cm]{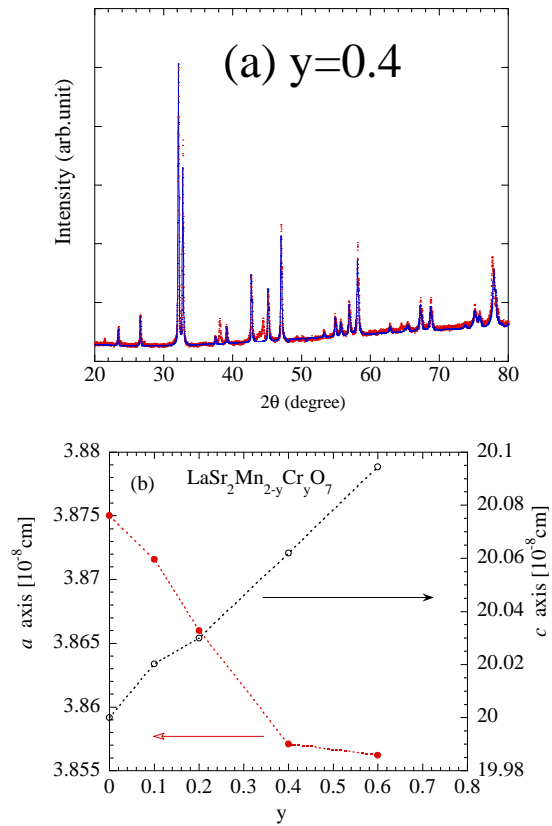}%
\caption{(Color online) (a) The x-ray powder diffraction pattern on the y=0.4 sample. Dots and a solid line are the observed and calculated intensities. The x-ray data are indexed in terms of (La,Sr)$_{3}$Mn$_{2}$O$_{7}$ 327
phase except for a small amount of impurity phase. (b) The lattice parameters calculated using the least squared fits as a function of Cr-content as listed in Table \ref{table1}. }
\label{xray}
\end{figure}%

The x-ray powder diffraction patterns were recorded for all samples on a RIGAKU diffractometer with  CuK $\alpha$ radiation as depicted in Fig.\ref{xray}. The x-ray data are indexed in terms of (La,Sr)$_{3}$Mn$_{2}$O$_{7}$ 327 phase except for a small amount of impurity phase, indicating a single phase of bilayered structure. The lattice parameters calculated using the least squared fits are listed in Table \ \ref{table1} as a function of Cr-content.
\begin{table}
\caption{\label{table1}  The lattice parameters, $a$ and $c$ , A-type AFM transition temperature $T_{N}$, spin-glass like transition temperature $T_{SG}$.
The $T_{N}$ is determined from a local maximum at higher $T$ in ZFC data while $T_{SG}$ is defined from the prominent peak  located at low-$T$. The lattice parameters of singlecrystalline LaSr$_{2}$Mn$_{2}$O$_{7}$ are taken from ref.\cite{SU00}.  }
\begin{ruledtabular}
\begin{tabular}{ccccc}
Sample&$a$ &$c$ &$T_N$&$T_{SG}$ \\
$y$& (\AA)& (\AA)&(K)&(K)\\
\hline
0&3.8790&19.996&210 &  \\
0.1&3.8716&20.020&175 & \\
0.2&3.8660&20.030&130 &   \\
0.4&3.8571&20.062& &38 \\
0.6&3.8562&20.094& &34.5 \\
\end{tabular}
\end{ruledtabular}
\end{table}

Magnetic  measurements as a function of temperature were carried out using a SQUID magnetometer in  both zero-field-cooled (ZFC) and field-cooled (FC) scans.  The magnetic relaxation was measured as follows;  First, the sample was cooled down to the respective temperatures in a zero field and then the applied field was held for 5 minutes. Finally, just after the field was switched off,  remanent magnetization data were recorded as a function of time.  Electrical resistivity was measured by a conventional four-probe technique.  Magnetoresistance measurements were performed at National Institute for Materials Science.  Here, an electric current supplied was parallel to the direction of the external field.  The thermal conductivity was measured using a conventional heat-flow method. The thermoelectric power $S$ (=$dV/dT$) was determined both from a temperature gradient and thermoelectric voltage, $dT$ and $dV$, which are generated from a thermal current  in the longitudinal direction of samples.    

\section{RESULTS  AND  DISCUSSION}
\subsection{Magnetic property}
\begin{figure}[ht]
\includegraphics[width=8cm]{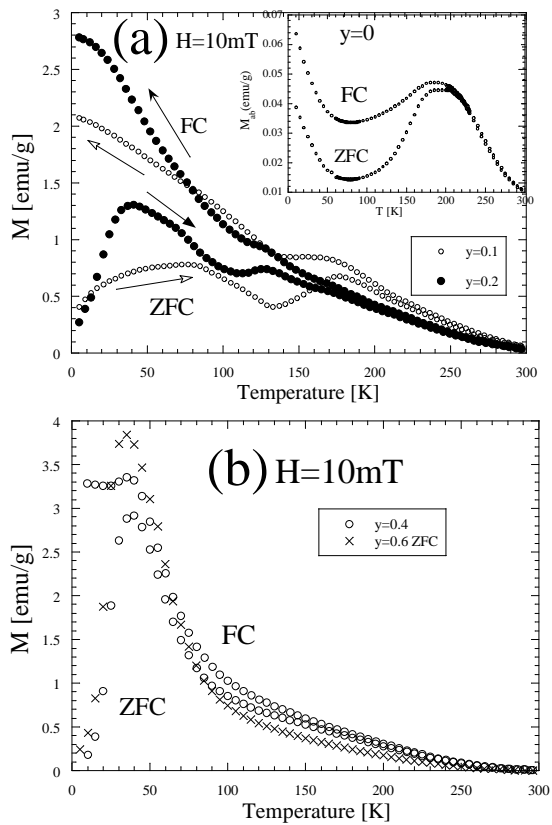}%
\caption{ ZFC and FC temperature dependences of the magnetization in polycrystalline 
LaSr$_{2}$Mn$_{2-y}$Cr$_{y}$O$_{7}$ ($y$=0.1,0.2,0.4 and 0.6),  measured at 10 mT . 
For comparison, the $ab$-plane magnetization data of parent crystal LaSr$_{2}$Mn$_{2}$O$_{7}$ are presented in the inset of(a).  }
\label{MT1}
\end{figure}%
\begin{figure}[ht]
\includegraphics[width=8cm]{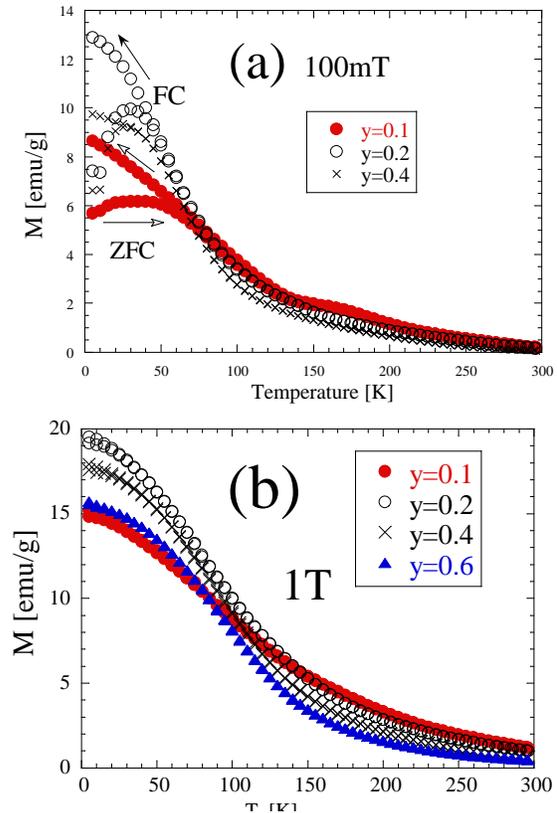}%
\caption{ (Color online) ZFC and FC temperature dependences of the magnetization in polycrystalline LaSr$_{2}$Mn$_{2-y}$Cr$_{y}$O$_{7}$ ($y$=0.1,0.2 and 0.4),  measured in a field of (a) 100mT and (b)  1 T. }
\label{MT2}
\end{figure}%

\begin{figure}[ht]
\includegraphics[width=8cm]{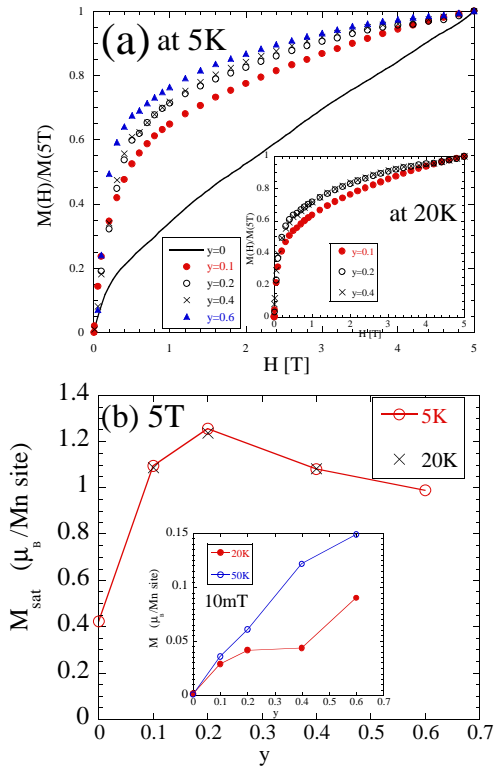}%
\caption{(Color online) (a) Field dependence of the magnetization at 5K in  LaSr$_{2}$Mn$_{2-y}$Cr$_{y}$O$_{7} $($y$= 0.1,0.2,0.4 and 0.6). The $M(H)$ data at 20K are also given in the inset of (a).
(b) the saturated magnetization at 5T as a function of Cr-content. For comparison, the low-field magnetization is plotted as a function of Cr-content in the inset of (b).}
\label{MH}
\end{figure}%
First, let us show in Fig. \ref{MT1}  the ZFC and FC temperature dependences of the magnetization in 
polycrystalline LaSr$_{2}$Mn$_{2-y}$Cr$_{y}$O$_{7}$ ($y$= 0.1,0.2,0.4 and 0.6),  measured at 10 mT . 
For comparison, the $ab$-plane magnetization data of parent crystal LaSr$_{2}$Mn$_{2}$O$_{7}$ are presented
 in the inset of Fig. \ref{MT1}  \cite{MMA03}.    Upon cooling the Cr-free sample, a broad  maximum in M$_{ab}$ is 
observed near about 210K, associated with the A-type AFM transition \cite{SU00,KI98}.   Cr-doping strongly suppresses Neel temperature $T_{N}$, from 210 K at $y$=0, through 175K at $y$=0.1, down to 130K at $y$=0.2  and  such a magnetic anomaly finally disappears for the $y$=0.4 and 0.6 samples.  The $T_{N}$ is determined from a local maximum at higher temperatures in ZFC data. In the A-type AFM structure,  FM spins lying in $ab$-plane of respective MnO$_{2}$ single layer are antiferromagnetically coupled  along the$ c$-axis within a MnO$_{2}$ double layer .  We expect that  a  partial substitution of Cr$^{3+}$ for Mn$^{3+}$ sites causes  $d_{x^2-y^2}$ orbital deficiencies of $e_g$-electron and weakens a AFM coupling working between respective single layers, resulting in an observed drop of $T_{N}$.   Instead, a low-$T$ peak in ZFC scan rapidly grows with Cr-doping, accompanied by a hysteresis region surrounded between ZFC and FC curves. At further low temperatures,  the ZFC magnetization of $y$=0.2-0.6  shows a steep decrease, indicating the freezing of magnetic moments \cite{DH01,AP01}.  These findings are reminiscent of  magnetic behaviors of a standard spin-glass system due to a magnetic frustration between ferromagnetic and antiferromagnetic interactions \cite{MY93}. A characteristic temperature  where the prominent peak in ZFC scan is located at low-$T$  is defined as $T_{SG}$ for the y=0.4 and0.6 samples at 10mT. 
In addition, the temperature variation of magnetization polycrystalline LaSr$_{2}$Mn$_{2-y}$Cr$_{y}$O$_{7}$ ($y$= 0.1,0.2 and 0.4 ) both in  0.1 T and 1T is shown in Fig. \ref{MT2}.   At 0.1T,  a history effect between ZFC and FC scans remains visible at lower-$T$. However, at relatively high field of 1T, the irreversibility in magnetization curves is strongly suppressed and a ferromagnetic-like behavior appears at low temperatures. These tendencies depending on the applied fields are never observed in a conventional spin-glass system.

Next, we examine the field dependence of low-$T$ magnetization in  LaSr$_{2}$Mn$_{2-y}$Cr$_{y}$O$_{7} $($y$=0.1,0.2,0.4 and 0.6) (Fig. \ref{MH} (a)).  The $ab$-plane magnetization of the Cr-free crystal shows a linear dependence on the field, in association with an AFM spin canting induced by the external field \cite{KI98}. On the other hand, in Cr-doped samples, $M-H$ curves rapidly  rise  at low fields and then tend to saturate up to a maximum field of 5T,indicating the development of ferromagnetic states. Upon increasing Cr-doping, the initial M shows a steeper rise.
Let us show in Fig. \ref{MH} (b) the saturated magnetization at 5T plotted as a function of Cr-content. We notice that the saturated magnetic moment $M_{sat}$ is almost independent of Cr-impurities in strong contrast to Cr-substitution effect on low-field magnetization in the inset of Fig. \ref{MH}(b).  It is true that Cr-impurity induces ferromagnetic moment from the inset of Fig. \ref{MH} (b), but the volume fraction of FM phase at 5T is almost insensitive of Cr content.  The value of  $M_{sat}$(5T) is converged within 30 to 35 percents of full ferromagnetic moment. ($M_{full}$=3.4 $\mu_{B}$ at y=0.2 and  $M_{full}$=3.2 $\mu_{B}$ at y=0.6.) 
We give some comments on the apparent disagreement in Cr-substitution effect between low and high field magnetic properties.  A partial substitution of Cr$^{3+}$ ion for Mn$^{3+}$ suppresses not only AFM coupling between single MnO$_{2}$  layers but also destroys FM double-exchange interaction between Mn$^{3+}$ and Mn$^{4+}$ ions  within the MnO$_{2}$  layer.  It is expected that the addition of Cr$^{3+}$ ions causes a suppression of FM region mediated by DE interaction through removing Mn$^{3+}$ ions.  
On the other hand, the low field data support the occurrence of  the ferromagnetic moment induced by Cr substitution. Following the Kanamori-Goodenough rules,  the superexchange(SE) interaction between Cr$^{3+}$ ($t_{2g}^3e_{g}^0$) and Mn$^{3+}$ ($t_{2g}^3e_{g}^1$) ions is ferromagnetic while the SE interaction between Cr$^{3+}$ and Mn$^{4+}$ ($t_{2g}^3e_{g}^0$))becomes antiferromagnetic \cite{GO63}.  
The annihilation of the Mn$^{3+}$ - Mn$^{4+}$ FM pairs is compensated by the creation of the Cr$^{3+}$ - Mn$^{3+}$ FM pairs accompanied by the Cr$^{3+}$ - Mn$^{4+}$ AFM pairs. In other words, the DE driven FM regions are partially replaced by the SE driven FM regions  with increasing the Cr ions, keeping the total FM fraction. 
The FM double-exchange interaction between Mn$^{3+}$ and Cr$^{3+}$ is not possible in our samples because the occurrence of FM moment by Cr-doping accompanies no metallic property as discussed later in the Cr-doping effect on resistivity.
At high fields, the phase separation between the field-induced FM phase and AFM second phase is probably realized at the level of clusters on the basis of  the competition between FM and AFM interaction. 
\begin{figure}[ht]
\includegraphics[width=8cm]{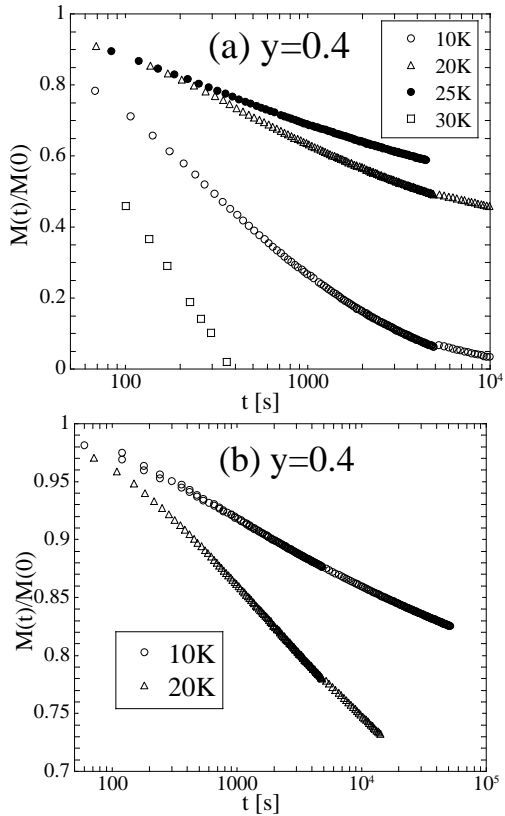}%
\caption{ Remanent magnetization data of the $y$=0.4 sample as a function of time, just after holding an applied field $H_{a}$ for 5 minutes and then switching it off.  (a) $H_{a}$=10mT and (b) $H_{a}$=100 mT.  }
\label{Relax}
\end{figure}%
Next, we carried out the magnetic relaxation of the $y$=0.4 sample  in order to examine the glassy state below $T_{SG}$.  In Fig.\ref{Relax}, we show the remanent magnetization data  of the $y$=0.4 sample as a function of time,  just after  holding an applied field for 5 minutes and then switching it off.  At 10mT,  the magnetization  relaxes faster at lower-$T$,  in contrast with the $M(t)$ data at 100mT.  However, at 1T  no slow relaxation in $M$ is observed, which is consistent with no history effect in ZFC and FC scans.   The slow decay of remanent magnetization curves indicates that a difference in free energy between the present excited and ground states is quite small in comparison with thermal energy and the system remains stable in various excited states \cite{MA01,DE02, GO01}.  Thus, a relatively fast relaxation of remanent $M$ at 10K in 10mT scan leads to a larger difference of energy  barrier between the ground and excited states than in the case of 100mT at the same temperature.  The metastable state excited by the lower field is probably related to the degree of a magnetic frustration between  AFM and FM clusters and/or the spatial  distribution of frustrated clusters.  Furthermore, the coexistence of frustrated clusters and ferromagnetic clusters  plays a crucial  role  in  the magnetic  relaxation  in 100mT.  FM spins and/or FM domain walls  are pinned  on the lattice defect sites like oxygen vacancy, giving a longer relaxation time.  

\subsection{Electrical  transport property }
\begin{figure}[ht]
\includegraphics[width=8cm,keepaspectratio]{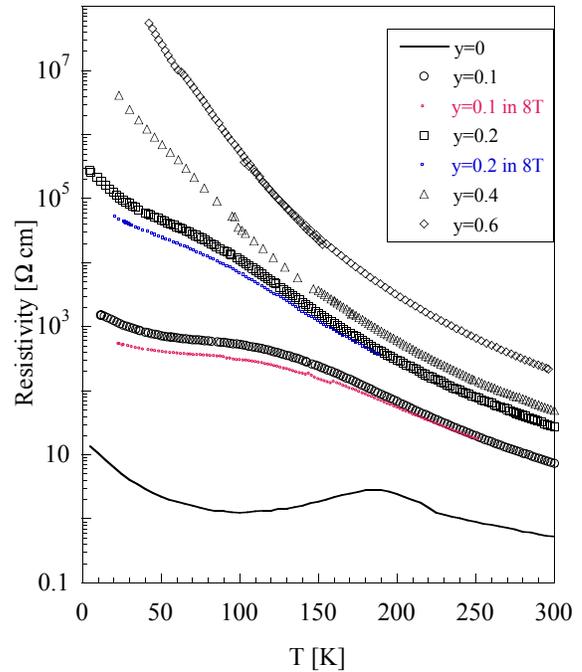}%
\caption{ (Color online) Temperature dependence of the electrical resistivity in polycrystalline LaSr$_{2}$Mn$_{2-y}$Cr$_{y}$O$_{7}$ ($y$=0.1,0.2,0.4 and 0.6). For comparison, the resistivity data of single crystalline LaSr$_{2}$Mn$_{2}$O$_{7}$ are also presented. }
\label{RT}
\end{figure}%
\begin{figure}[ht]
\includegraphics[width=8cm]{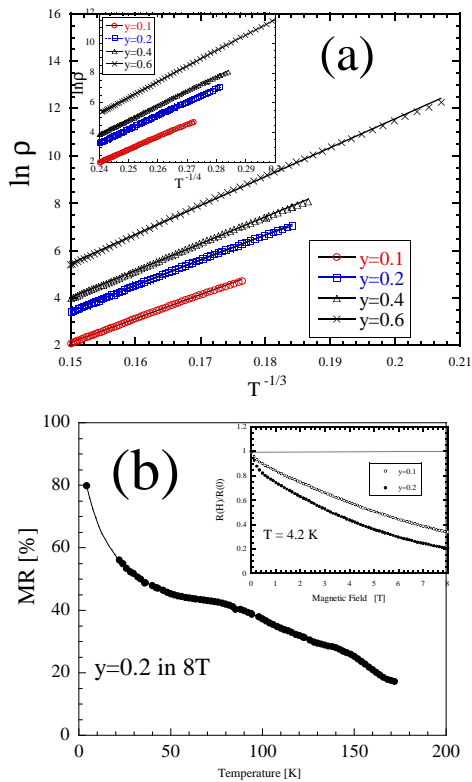}%
\caption{(Color online) (a) a semilog plot of $\rho$ versus $T^{-p}$ with $p=1/3$ for LaSr$_{2}$Mn$_{2-y}$Cr$_{y}$O$_{7}$ ($y$=0.1,0.2,0.4 and 0.6). 
The solid lines correspond to fits by Mott's VRH model. The inset also presents a semilog plot of $\rho$ versus $T^{-p}$ with $p=1/4$ for 3D VRH. 
With increasing Cr-doping level, the VRH regime is extended as listed in Table\ \ref{table2}. 
(b) Magnetoresistance effect of the $y$=0.2 sample as a function of temperature at 8T. In the inset , the MR of $y$= 0.1 and 0.2 samples at 4.2K is plotted as a function of field up to 8T. }
\label{RT2}
\end{figure}%

Figure \ref{RT} displays the temperature dependence of the electrical resistivity in polycrystalline LaSr$_{2}$Mn$_{2-y}$Cr$_{y}$O$_{7}$ ($y$=0.1,0.2,0.4 and 0.6). For comparison, the resistivity data of parent crystal LaSr$_{2}$Mn$_{2}$O$_{7}$ are also presented in Fig.\ref{RT} .  The value of $\rho$ at lower $T$ exhibits a rapid increase by about four orders of magnitude,  from $\sim$ 10$^2$ $\Omega$cm at $y$=0.1 up to $10^6$ $\Omega$cm at $y$=0.4.  Cr-doping strongly enhances an insulating behavior over a wide range of temperature because conduction paths are partially destroyed by  $d_{x^2-y^2}$ orbital deficiencies of $e_g$-electron. Our data exclude in this system a possibility of the global double-exchange interaction between Mn$^{3+}$ and Cr$^{3+}$ ions,  giving a metallic property \cite{SUN01}.    
 In particular, for the $y$=0-0.2 sample, the rapid rise in $\rho(T)$ below 50K is close to carrier localization effect  due to a suppression of carrier hopping between single layers because at lower-$T$  orbital fluctuation of $d_{x^2-y^2}$  is gradually suppressed and motion of carriers are confined within respective single layer \cite{KI98}.  

\begin{table}
\caption{\label{table2}  The fitting parameters,  $\rho_{0}$  and $T_{0}$ for polycrystalline samples of  LaSr$_{2}$Mn$_{2-y}$Cr$_{y}$O$_{7}$ ($y$=0.1, 0.2, 0.4 and 0.6)  }
\begin{ruledtabular}
\begin{tabular}{cccccc}
&&2D VRH&&3D VRH&\\
Sample&VRH regime & $\rho_{0}$ &$T_{0}$&$\rho_{0}$&  $T_{0}$ \\
$y$&(K)&($\Omega$cm)&(K)& ($\Omega$cm)&(K)\\
\hline
0.1& $T>187 $ & $2.1\times10^{-6}$ & $1.0\times10^{6}$ & $8.7\times10^{-9}$  &$5.4\times10^{7}$  \\
0.2& $T>161 $ & $3.0\times10^{-6}$ & $1.2\times10^{6}$ & $8.1\times10^{-9}$  &$7.0\times10^{7}$  \\

0.4& $T>155 $ & $2.0\times10^{-6}$ & $1.5\times10^{6}$ & $3.7\times10^{-9}$  &$8.9\times10^{7}$  \\

0.6&  $T>113$ & $2.6\times10^{-6}$ & $1.8\times10^{6}$ & $2.7\times10^{-9}$  &$1.2\times10^{8}$  \\
\end{tabular}
\end{ruledtabular}
\end{table}

We try to analyze the $\rho(T)$ data of Cr-doped samples using the small-polaron hopping model and
 Mott's variable-range-hopping (VRH) model\cite{MOTT79}, to examine the conduction mechanism of bilayered manganites\cite{CH03}.  
According to Mott's VRH model, the temperature dependence of resistivities  is represented 
by $\rho(T)$=$\rho_{0}$exp[($T_{0}$/$T$)$^{p}$], where $\rho_{0}$ is a constant and $p=1/(d+1)$ with $d$
 being the dimensionality of the system. Mott's activation energy $T_{0}$ is proportional to $1/[N(E)\xi^{d} ]$, 
where $N(E)$ is the density of states at the Fermi level and $\xi$ is the localization length.
 On the other hand,  the adiabatic small-polaron model is described by $\rho(T)$=$\rho_{0}$$T$exp($E_{\rho}$/$kT$) , 
where $\rho_{0}$ is a constant and $E_{\rho }$ represents  the activation energy of small-polaron.  
For all samples with Cr-substitution, it is found that the VRH model gives a more reasonable fit to the experimental data over a wide range of temperatures, in comparison with the small-polaron model. In Fig.7(a), we present our results as a semilog plot of $\rho$ versus $T^{-p}$ with $p=1/3$ for 2D VRH,  while the inset of Fig.7(a) shows a semilog plot of $\rho$ versus $T^{-p}$ with $p=1/4$ for 3D VRH. Although it is hard to distinguish a $T^{-1/3}$ or $T^{-1/4}$ dependence of ln$\rho$,  we obtain a much better fit to Mott's VRH than to a VRH model with $p=1/2$ in the presence of Coulomb gap\cite{CH03,SH84}. The fitting parameters,  $\rho_{0}$  and $T_{0}$, for polycrystalline samples of  LaSr$_{2}$Mn$_{2-y}$Cr$_{y}$O$_{7}$ ($y$=0.1, 0.2, 0.4 and 0.6)  are listed in Table\ \ref{table2}.  With increasing Cr-content, the value of $T_{0}$ shows a monotonous increase for both 2D and 3D cases, indicating the decrease of  the localization length$\xi$. The localization effect  enhanced due to Cr-substitution is probably associated with orbital disorders in Mn-O-Mn networks   introduced by the removal of $e_g$-electron \cite{ZHANG04,NA04}.  

We give some comments on the doping effect of other trivalent metallic ions (Co$^{3+}$ and Al$^{3+}$) on the Mn sites of La$_{1}$Sr$_{2}$Mn$_{2}$O$_{7}$ \cite{ZHANG04,NA04}.  The 3$d$ electronic configuration of Co$^{3+}$ ion follows as; $t_{2g}^6e_{g}^0$ ($S$=0,low-spin state), $t_{2g}^5e_{g}^1$ ($S$=1,intermediate -spin state) and  $t_{2g}^4e_{g}^2$ ($S$=2,high-spin state). The Al$^{3+}$ ion is a non magnetic ion without  $d$-electrons.  With increasing Co$^{3+}$(or Al$^{3+}$) doping level, the A-type AFM temperature shifts to low temperatures and the magnitude of magnetization decreases over a wide range of temperatures. The decrease of $M$ implies a reduction of the net magnetic moments, which is consistent with low-spin state (S=0) of  Co$^{3+}$ or non magnetic ion of Al$^{3+}$. The latter tendency  is in strong contrast with the magnetic effect of Cr$^{3+}$ ($t_{2g}^3e_{g}^0$,S=3/2) doping on La$_{1}$Sr$_{2}$Mn$_{2}$O$_{7}$ although a suppression of Atype-AFM temperature is commonly  observed for Cr, Co and Al doping. 
On the other hands, the doping effects on electrical transport for Cr, Co and Al ions exhibit such common features as the enhanced insulating state due to orbital deficiencies following the VRH model. In particular, the Al substitution without d-electrons for Mn site products a more rapid increase in resistivities.

  Magnetoresistance(MR) effect of the $y$=0.2 sample as a function of temperature is depicted in Fig.\ref{RT2} (b),  where the negative MR is defined as -100*[$\rho$(8T)-$\rho$ (0T)]/$\rho$ (0T) . The value of giant MR increases from 25 \% at 150K  up to 80 \% at 4.2K with decreasing $T$.  The existence of  the field-induced FM clusters is probably related to the enhanced MR at low temperatures as we see from MT data in Fig.\ref{MT2}(b).
 In the inset of Fig. \ref{RT2} (b),  the MR of $y$= 0.1 and 0.2 samples is plotted as a function of field up to 8T.  Cr-doping also increases a low-$T$ MR from 45 \% at $y$=0.2   up to 80 \% at $y$=0.6 under a field of 8T at 40K. The Cr-doping induced orbital disorders  assist charge transfer along the $c$-axis  across respective single layer  of  MnO$_{2}$,  giving the enhanced MR effect.

\subsection{Thermal transport properties (Seebeck effect and thermal conductivity) }
\begin{figure}[ht]
\includegraphics[width=8cm]{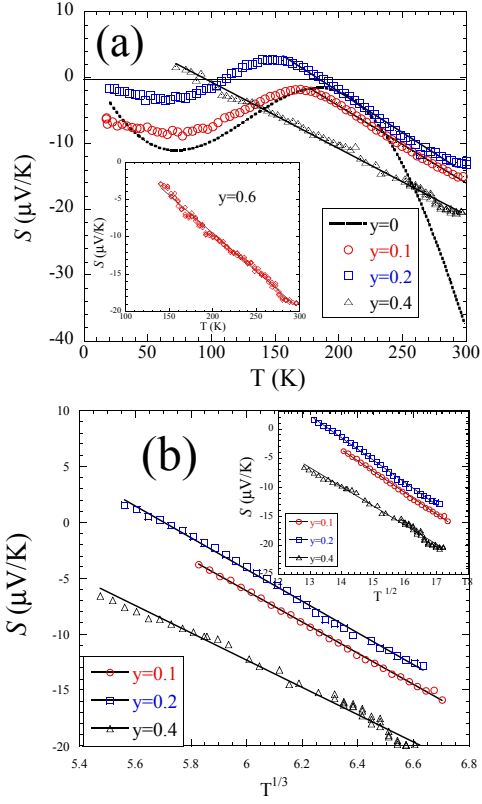}%
\caption{(Color online) (a) Temperature variation of Seebeck coefficient $S$  for LaSr$_{2}$Mn$_{2-y}$Cr$_{y}$O$_{7}$
 ($y$=0, 0.1,0.2 and 0.4). The solid lines correspond to $T$ linear fits. The inset represents the $S$ data of the $y$=0.6 sample with a linear fit. 
 (b) Seebeck coefficient $S(T)$  versus $T^{1/2}$  for LaSr$_{2}$Mn$_{2-y}$Cr$_{y}$O$_{7}$ 
($y$=0.1,0.2 and 0.4). The solid lines correspond to fits by the 2D VRH model. The inset presents  $S(T)$  versus $T^{1/3}$
 for the 3D VRH model. In the case of $y$=0.2, we have typical fitting parameters $A$=77 $\mu$V/K and $B$=14 $\mu$V/K$^{3/2}$
 for 2D VRH ($A$=52 $\mu$V/K and $B$=3.8 $\mu$V/K$^{4/3}$ for 3D VRH), where $S(T)=A-B T^{p}$. The $S(T)$ of
 the $y$=0.2 follows the VRH law for 169 K $\leq$ $T$ $\leq$ 300 K}
\label{ST}
\end{figure}%

Next, the temperature variation of Seebeck coefficient $S$  for the $y$=0.1-0.6 samples is displayed in Fig.\ref{ST}(a). 
For comparison,  the $S(T)$ data of single crystalline  LaSr$_{2}$Mn$_{2}$O$_{7}$ are cited \cite{MMA03}.
 For $y$=0-0.2, with decreasing $T$,  the value of $S(T)$ shows  a local maximum near the A-type AFM transition temperature $T_{N}$ and
 then a shallow minimum at lower $T$ is observed \cite{NA01}.  At lower $T$, Cr-doping gradually suppresses a local minimum of $S(T)$ from
 a negative  value at $y$=0 down to a small one at $y$=0.2 and finally at $y$=0.4 the local minimum in $S(T)$ disappears, 
giving a monotonous decrease over the observed temperature range. Now, let us try to analyze the $S(T)$ data  of Cr-doped samples using the extended  Mott's VRH model to Seebeck coefficients \cite{MOTT79,ZV91}.  For the 2D VRH case, the corresponding form is described by $S(T)\propto T^{p}$ with $p=1/3$ ($p=1/2$ for the 3D VRH case). In Fig.8(b), we present our results as a linear plot of $S(T)$versus $T^{1/2}$ for 2D VRH (in the inset, $S(T)$ versus $T^{1/3}$ for 3D VRH).  In a similar way, we obtain a much better fit of $S(T)$ data  to the VRH law than to the thermally activated $T$ dependence. Here, the Seebeck coefficient for a thermally activated case is expressed as  $S(T)$=$k/e$($E_{S}$/$kT$)+$S_{\infty}$, where $E_{S}$ is a thermal activation energy  and $S_{\infty}$ denotes Seebeck coefficient in the high temperature limit. In addition, the obvious differences among the $T$, $T^{1/2}$ and $T^{1/3}$ dependences we do not notice within our fitting procedures. In the random hopping system, the $T$-linear dependence of $S(T)$ is theoretically presented by Culter and Mott \cite{CU69}. The $T$-linear dependence of $S(T)$ in the insulating state is probably related to a random distribution of localized electronic states around Fermi level as reported in Seebeck coefficient of Li$_{1+x}$Ti$_{2-x}$ O$_{4}$ oxides ref.\cite{MA87}.

In a doped bilayer manganite with hole content $x$=0.4 , the high temperature behavior of $S(T)$  is well explained on the basis of  a model of Zener polarons, where a Zener polaron formed in the high-$T$ region occupy two manganese sites \cite{ZH98}.   It is true that  this model qualitatively  reproduces a negative sign in high- $T$  behavior  of single crystalline La$_{1}$Sr$_{2}$Mn$_{2}$O$_{7}$. However, for all polycrystalline samples with Cr-substitution it seems that the VRH conduction gives a reasonable fit to both resistivities and Seebeck coefficients.

\begin{figure}[ht]
\includegraphics[width=8cm]{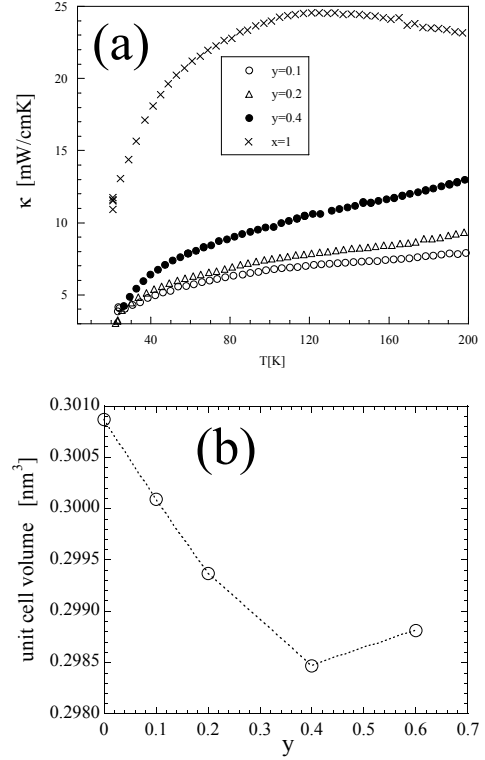}%
\caption{(a)Temperature dependence of thermal conductivity for
 LaSr$_{2}$Mn$_{2-y}$Cr$_{y}$O$_{7}$ ($y$=0.1,0.2, and 0.4 ). For comparison, the $\kappa$ data of polycrystalline Sr$_{3}$Mn$_{2}$O$_{7}$ ($x$=1)  is presented. (b) the unit-cell volume of
 LaSr$_{2}$Mn$_{2-y}$Cr$_{y}$O$_{7}$ as a function of  Cr-content. }
\label{KT}
\end{figure}%

Finally, let us show in Fig.\ref{KT} the thermal conductivity  of Cr-doped LaSr$_{2}$Mn$_{2-y}$Cr$_{y}$O$_{7}$ ($y$=0.1,0.2, and 0.4 ) as a function of temperature.  For comparison, the $\kappa$ data of polycrystalline Sr$_{3}$Mn$_{2}$O$_{7}$ ($x$=1)  is presented \cite{MA03}.  First of all, thermal carries are  phonons since the electron component is estimated to be negligible from the resistivity data using the Widemann-Franz law.  The phonon thermal conduction gradually increases with Cr-doping, which seems to be an unusual behavior because  the introduction of Cr-impurity ions would disturb phonon conduction. However, this anomalous finding is reasonably resolved through clarifying a close relationship between phonon conduction and local lattice distortion of MnO$_{6}$  due to Jahn-Teller effect. In our previous work on thermal conductivity in bilayered manganite single crystals, it has been  made clear that  the phonon conduction in the insulating state is scattered by local lattice distortions of Mn$^{3+}$ O$_{6}$ but the metallic state realized by lowering of $T$ or by the applied field yields a upturn in $\kappa$ below $T_{C}$ or giant magnetothermal effect\cite{MA03}.   This enhanced phonon conduction arises from a suppression of Mn$^{3+}$ O$_{6}$ local distortions due to a screening effect of itinarent carriers.  Cr-substitution for Mn$^{3+}$  sites removes $d_{x^2-y^2}$  orbitals of $e_g$-electron ,  resulting in a Cr$^{3+}$ O$_{6}$ octahedron without local JT effect. In other words, Cr-doping effect on lattices causes a suppression of local lattice distortion through the introduction of  JT  inactive ions, giving an increase in phonon conduction. Surely, the $ \kappa (T)$ of polycrystalline Sr$_{3}$Mn$_{2}$O$_{7}$($x$=1) shows a typical phonon conduction, whose behavior is free from JT distortion of Mn$^{3+}$ O$_{6}$.  In addition,  the Cr-doping dependence of $a-$ and $c-$ axis lattice parameters in Table\ref{table1} reveals the volume shrinkage of the unit cell  with increasing Cr-content as shown in Fig.8 (b).  We note that the lattice constant of $y$=0.6 is influenced by a small amount of the impurity phase. This volume effect  is associated with a number of  deficiencies of $d_{x^2-y^2}$  orbitals of $e_g$-electron,  which is quite consistent with the preceding discussion on the close relationship between the lattice distortion and phonon conduction.

 \section{SUMMARY}
 
 We have carried out magnetic, electrical, Seebeck effect and thermal conductivity measurements of LaSr$_{2}$Mn$_{2-y}$Cr$_{y}$O$_{7}$ polycrystalline samples ($y$=0.1, 0.2, 0.4 and 0.6).  
The Cr$^{3+}$ substitution for Mn$^{3+}$ sites produces a monotonic shrink of $a$($b$)-axis in contrast with a gradual elongation of $c$-axis in association with 
 a removal  of  $d_{x^2-y^2}$ orbital  of  $e_g$-electron. 
For Cr-doped samples, a glassy behavior appears accompanied by both a collapse of the A-type antiferromagnetic property and the development of ferromagnetic clusters.
At high fields, the irreversibility in magnetization curves disappears and the saturated magnetic moment induced by the applied field reaches 30 to 35 percents of full ferromagnetic moment at 5T for all Cr doped samples.  This finding strongly suggests the presence of a phase separation between FM and second phases at the level of clusters, which originates from the frustration between FM and AFM interactions.
The electrical transport for Cr-doped samples strongly enhances an insulating property  over the wide range of temperature because conduction paths are partially destroyed by  $d_{x^2-y^2}$ orbital deficiencies of $e_g$-electron. 
At lower $T$, Cr-doping gradually suppresses a local minimum of $S(T)$ from a relatively large value at $y$=0 down to a positively small one at $y$=0.4  in striking contrast to the more enhanced low-$T$ resistivity data. For all polycrystalline samples with Cr-substitution it seems that the VRH conduction gives a reasonable fit to both resistivities and Seebeck coefficients.   
The phonon thermal conduction gradually increases with increasing Cr content, which is contrast to a typical impurity effect on thermal conductivity. We  propose that the increase in the phonon thermal conduction results from a suppression of local lattice distortion through the introduction of Jahn-Teller inactive ion of Cr$^{3+}$.

%
\begin{acknowledgments}

This work was partially supported by a Grand-in-Aid for Scientific Research from the Ministry of Education, Science and Culture, Japan. The authors thank Mr. H.Noto and Dr. S. Ueda for their technical supports. 
\end{acknowledgments}


\end{document}